\documentstyle[amssymb,aps,preprint]{revtex}

\begin{document}
\title{Decay Rate of A Wannier Exciton in Low Dimensional Systems}
\author{Yueh-Nan Chen and Der-San Chuu\thanks{%
corresponding author: e-mail: dschuu@cc.nctu.edu.tw; Fax: 886-3-5725230;
Tel: 886-3-5712121-56105 }}
\address{Department of Electrophysics, National Chiao Tung University,\\
Hsinchu 30050, Taiwan}
\date{\today}
\maketitle

\begin{abstract}
The superradiant decay rate of Wannier exciton in one dimensional system is
studied. The crossover behavior from 1D chain to 2D film is also examined.
It is found that the decay rate shows oscillatory dependence on channel
width $L$. When the quasi 1-D channel is embeded with planar microcavities,
it is shown that the dark mode exciton can be examined experimentally.
\end{abstract}

\newpage

\bigskip

Since Dicke\cite{1} pointed out the concept of superradiance, the coherent
effect for spontaneous radiation of various systems has attractted extensive
interest both theoretically and experimentally\cite{2,3}. The coherent
radiation phenomena for the atomic system was intensively investigated in
the late sixties\cite{4,5,6}. One of the limiting case of superradiance is
the exciton-polariton state in solid state physics. When a Frenkel exciton
couples to the radiation field in a small system which contains $N$ lattice
points, it represents one excited atom in some site and $N-1$ unexcited
atoms in the others. According to Dicke's theory, the decay rate of the
system will be enhanced by the factor of $N$. But as it was well known in a
3-D bulk crystal,{\it \ }the excitons will couple with photons to form
polaritons--the eigenstate of the combined system consisting of the crystal
and the radiation field which does not decay radiatively.\cite{7} What makes
the excitons trapped in the bulk crystal is the conservation of crystal
momentum. If one considers a linear chain or a thin film, the exciton can
undergo radiative decay as a result of the broken crystal symmetry. The
decay rate of the exciton is enhanced by a factor of $\lambda /d$ in a
linear chain\cite{8} and ($\lambda /d)^{2}$ for 2D exciton-polariton\cite
{9,10}, where $\lambda $ is the wave length of emitted photon and $d$ is the
lattice constant of the linear chain or the thin film.

Lots of investigations on the radiative linewidth of the excitons have been
performed. First observation of superradiant short lifetime have been
performed by Ya. Aaviksoo {\em et al}.\cite{11} on surface states of the
anthracene crystal. Latter, B. Deveaud {\em et al}.\cite{12} measured the
radiative lifetime of free excitons in GaAs quantum wells and observed the
enhanced radiative recombination of the excitons. Hanamura\cite{13}
investigated theoretically the radiative decay rate of quantum dot and
quantum well. The obtained results are in agreement with that of Lee and
Liu's\cite{10} prediction for thin films. Knoester\cite{14} obtained the
dispersion relation of Frankel excitons of quantum slab. The oscillating
dependence of the radiative width of the excitonlike polaritons with the
lowest energy on the crystal thickness was found. Recently, G. Bj\"{o}rk 
{\em et al}.\cite{15} examined the relationship between atomic and excitonic
superradiance in thin and thick slab geometries. They demonstrated that
superradiance can be treated by a unified formalism for atoms, Frenkel
excitons, and Wannier excitons. In V. M. Agranovich {\em et al}.'s work\cite
{16}, a detailed microscopic study of Frenkel exciton-polariton in crystal
slabs of arbitrary thickness was performed.

For lower dimensional systems, A. L. Ivanov and H. Haug\cite{17} predicted
the existence of exciton crystal, which favors coherent emission in the form
of superradiance, in quantum wires. Y. Manabe {\em et al}.\cite{18}
considered the superradiance of interacting Frenkel excitons in a linear
chain. Recently, with the advances of the modern fabrication technology, it
has become possible to fabricate the planar microcavities incorporating
quantum wires\cite{19}. This makes it interesting to study this problem more
carefully. In this paper, we will investigate the radiative decay of the
Wannier exciton in a quantum channel and the crossover behavior from 1D
chain to 2D film. Application to real semiconductor microcavity will also be
considered.

Let us first consider a linear chain with lattice spacing $d$. The state of
the Wannier exciton can be specified as $\left| k_{z},n\right\rangle $,
where $k_{z}$ is the exciton wave number in the chain direction and $n$ is
the quantum number for internal structure of the exciton. The Hamiltonian
for the exciton is

\begin{equation}
H_{ex}=\sum_{k_{z}n}E_{k_{z}n}c_{k_{z}n}^{\dagger }c_{k_{z}n},
\end{equation}
where $c_{k_{z}n}^{\dagger }$ and $c_{k_{z}n}$ are the creation and
destruction operators of the exciton, respectively. The Hamiltonian of free
photon is

\begin{equation}
H_{ph}=\sum_{{\bf q}^{\prime }k_{z}^{\prime }\lambda }\hbar c(q^{\prime
2}+k_{z}^{\prime 2})^{1/2}b_{{\bf q}^{\prime }k_{z}^{\prime }\lambda
}^{\dagger }b_{{\bf q}^{\prime }k_{z}^{\prime }\lambda },
\end{equation}
where $b_{{\bf q}^{\prime }k_{z}^{\prime }\lambda }^{\dagger }$ and $b_{{\bf %
q}^{\prime }k_{z}^{\prime }\lambda }$ are, respectively, the creation and
destruction operators of the photon, and $\lambda $ runs through the indices
of $x,y,$ and $z$. The wave vector ${\bf k}^{\prime }$\hspace{0.06in}of the
photon were separated into two parts: $k_{z}^{\prime }$ is the parallel
component of ${\bf k}^{\prime }$ along the linear chain such that $k^{\prime
2}=q^{\prime 2}+k_{z}^{\prime 2}$.

The interaction between the exciton and the photon can be expressed as

\begin{eqnarray}
H^{\prime } &=&\sum_{i}\sum_{{\bf q}^{\prime }k_{z}^{\prime }\lambda }\frac{e%
}{mc}\sqrt{\frac{2\pi \hbar c}{(q^{\prime 2}+k_{z}^{\prime 2})^{1/2}v}} 
\nonumber \\
&&\times [b_{{\bf q}^{\prime }k_{z}^{\prime }\lambda }^{\dagger }\exp
(ik_{z}^{\prime }\tau _{i})+b_{{\bf q}^{\prime }k_{z}^{\prime }\lambda }\exp
(-ik_{z}^{\prime }\tau _{i})]({\bf \epsilon }_{{\bf q}^{\prime
}k_{z}^{\prime }\lambda }\cdot {\bf p}_{i}),
\end{eqnarray}
where $m$ is the electron mass, ${\bf \tau }_{i}$ is a position vector of
the electron $i$ in the linear chain, ${\bf p}_{i}$ is the corresponding
momentum of the electron $i$\ operator, and ${\bf \epsilon }_{{\bf q}%
^{\prime }k_{z}^{\prime }\lambda }$ is the polarization vector of the photon.

The essential quantity involved is the matrix element of $H^{\prime }$
between the ground state $\left| G\right\rangle $ and the Wannier exciton
state $\left| k_{z},n\right\rangle $. We know that the interaction matrix
elements of $H^{\prime }$ can be written as

\begin{equation}
\left\langle k_{z},n\right| H^{\prime }\left| G\right\rangle =\sum_{l,\rho
}\left\langle c,l+\rho ;v,l\right| U_{k_{z}n}^{\ast }(l,\rho )H^{\prime
}\left| G\right\rangle ,
\end{equation}
because the Wannier exciton state can be expressed as

\begin{equation}
\left| k_{z},n\right\rangle =\sum_{l,\rho }U_{k_{z}n}^{*}(l,\rho )\left|
c,l+\rho ;v,l\right\rangle ,
\end{equation}
in which the excited state $\left| c,l+\rho ;v,l\right\rangle $ is defined as

\begin{equation}
\left| c,l+\rho ;v,l\right\rangle =a_{c,l+\rho }^{\dagger }a_{v,l}\left|
G\right\rangle ,
\end{equation}
where $a_{l+\rho }^{\dagger }(a_{v,l})$ is the creation (destruction)
operator of an electron in the conduction band $\left( c\right) $ (valence
band $\left( v\right) $) band at lattice site $l+\rho (l).$ The expansion
coefficient $U_{k_{z}n}^{*}(l,\rho )$ is the exciton wave function in the
linear chain:

\begin{equation}
U_{k_{z}n}^{*}(l,\rho )=\frac{1}{\sqrt{N^{\prime }}}e^{ik_{z}r_{c}}F_{n}(%
\rho ),
\end{equation}
where the coefficient $1/\sqrt{N^{\prime }}$ is for the normalization of the
state $\left| k_{z},n\right\rangle ,$ and $F_{n}(\rho )$ and $r_{c}$
represent the one-dimensional hydrogenic wavefunction and the center of mass
of the exciton, respectively.

After summing over $l$ in eq.(4), we have

\begin{eqnarray}
\left\langle k_{z},n\right| H^{\prime }\left| G\right\rangle &=&\sum_{{\bf q}%
^{\prime }\lambda }\frac{e}{mc}\sqrt{\frac{2\pi \hbar c}{(q^{\prime
2}+k_{z}^{2})^{1/2}v}}  \nonumber \\
&&\times [b_{{\bf q}^{\prime }k_{z}\lambda }({\bf \epsilon }_{{\bf q}%
^{\prime }k_{z}\lambda }\cdot {\bf A}_{k_{z}n})+b_{-{\bf q}^{\prime
}k_{z}\lambda }^{\dagger }({\bf \epsilon }_{-{\bf q}^{\prime }k_{z}\lambda
}\cdot {\bf A}_{k_{z}n})],
\end{eqnarray}
where

\begin{eqnarray}
{\bf A}_{k_{z}n} &=&\sqrt{N^{\prime }}\sum_{\rho }F_{n}(\rho )\int d{\bf %
\tau }w_{c}({\bf \tau }-\rho )  \nonumber \\
&&\times \exp (ik_{z}(\tau -\frac{m_{e}^{*}\rho }{m_{e}^{*}+m_{h}^{*}}%
))(-i\hbar {\bf \nabla })w_{v}({\bf \tau }),
\end{eqnarray}
, $w_{c}({\bf \tau })$ and $w_{v}({\bf \tau })$ are, respectively, the
Wannier functions for the conduction band and the valence band at site 0,
and $m_{e}^{*}$ and $m_{h}^{*}$ are, respectively, the effective masses of
the electron and hole. Hence the interaction between the exciton and the
photon (in the resonance approximation) can be written in the form 
\begin{equation}
H^{\prime }=\sum_{k_{z}n}\sum_{{\bf q}^{\prime }\lambda }D_{{\bf q}^{\prime
}k_{z}n}b_{k_{z}{\bf q}^{\prime }\lambda }c_{k_{z}n}^{\dagger }+{\bf h.c.,}
\end{equation}
where

\begin{equation}
D_{{\bf q}^{\prime }k_{z}n}=\frac{e}{mc}\sqrt{\frac{2\pi \hbar c}{(q^{\prime
2}+k_{z}^{\prime 2})^{1/2}v}}{\bf \epsilon }_{{\bf q}^{\prime }k_{z}\lambda
}\cdot {\bf A}_{k_{z}n}.
\end{equation}

Now, we assume that at time $t=0$ one of the Wannier excitons is in the mode 
$k_{z},n.$ For time $t>0$,\hspace{0.06in}the state $\left| \psi
(t)\right\rangle $\hspace{0.06in}can be written as

\begin{equation}
\left| \psi (t)\right\rangle =f_{0}(t)\left| k_{z},n;0\right\rangle +\sum_{%
{\bf q}^{\prime }}f_{G;{\bf q}^{\prime }k_{z}}(t)\left| G;{\bf q}^{\prime
}k_{z}\right\rangle ,
\end{equation}
where $\left| k_{z},n\right\rangle $ is the state with a Wannier exciton in
the mode $k_{z},n$ in the linear chain, $\left| G;{\bf q}^{\prime
}k_{z}\right\rangle $ represents the state in which the electron-hole pair
recombines and a photon in the mode ${\bf q}^{\prime }{\bf ,}k_{z}$ is
created, and $f_{0}(t)$ and $f_{G;{\bf q}^{\prime }k_{z}}(t)$ are,
respectively, the probability amplitudes of the state $\left|
k_{z},n\right\rangle $ and $\left| G;{\bf q}^{\prime }k_{z}\right\rangle $.

By the method of Heitler and Ma in the resonance approximation, the
probability amplitude $f_{0}(t)$ can be expressed as\cite{10}

\begin{equation}
f_{0}(t)=\exp (-i\Omega _{k_{z}n}t-\frac{1}{2}\gamma _{_{k_{z}n}}t),
\end{equation}
where

\begin{equation}
\gamma _{_{k_{z}n}}=2\pi \sum_{{\bf q}^{\prime }\lambda }\left| D_{{\bf q}%
^{\prime }k_{z}n}\right| ^{2}\delta (\omega _{{\bf q}^{\prime }k_{z}n})
\end{equation}
and

\begin{equation}
\Omega _{k_{z}n}={\cal P}\sum_{{\bf q}^{\prime }}\frac{\left| D_{{\bf q}%
^{\prime }k_{z}n}\right| ^{2}}{\omega _{{\bf q}^{\prime }k_{z}n}}
\end{equation}
with $\omega _{{\bf q}^{\prime }k_{z}n}=E_{k_{z}n}/\hbar -c\sqrt{q^{\prime
2}+k_{z}^{2}}.$ Here $\gamma _{_{k_{z}n}}$ and $\Omega _{k_{z}n}$ are,
respectively, the decay rate and frequency shift of the exciton. And ${\cal P%
}$ means the principal value of the integral.

The Wannier exciton decay rate in the optical region be calculated
straightforwardly and is given by

\begin{equation}
\gamma _{k_{z}n}=\left\{ 
\begin{array}{c}
\frac{3\pi }{2k_{0}d}\gamma _{0}\sum\limits_{\lambda }\frac{\left| {\bf %
\epsilon }_{k_{z}\lambda }\cdot \chi _{n}\right| ^{2}}{\left| \chi
_{n}\right| ^{2}},\text{ }k_{z}<k_{0} \\ 
0,\text{ otherwise \ \ \ \ \ \ \ \ \ \ \ \ \ \ \ \ \ \ \ \ \ \ \ \ \ }
\end{array}
\right. ,
\end{equation}
where $k_{0}=E_{k_{z}n}/\hbar $,

\begin{equation}
{\bf \chi }_{n}=\sum_{\rho }F_{n}^{*}(\rho )\int d{\bf \tau }w_{c}({\bf \tau 
}-\rho )(-i\hbar {\bf \nabla })w_{v}({\bf \tau }),
\end{equation}
and

\begin{equation}
\gamma _{0}=\frac{4e^{2}\hbar k_{0}}{3m^{2}c^{2}}\left| {\bf \chi }%
_{n}\right| ^{2}.
\end{equation}

Here, ${\bf \chi }_{n}^{*}$ represents the effective dipole matrix element
for an electron jumping from the excited Wannier state in the conduction
band back to the hole state in the valence band, and $\gamma _{0}$ is the
decay rate of an isolated atom. We see from eq.(16) that $\gamma
_{_{k_{z}n}} $ is proportional to $1/(k_{0}d)$. This is just the
superradiance factor coming from the coherent contributions of atoms within
half a wavelength or so\cite{8,18,20}$.$

Now let us consider a quasi-1D channel with channel width $L=Nd$. The state
of the Wannier exciton can be specified as $\left| {\bf k},n,l\right\rangle $%
, where ${\bf k}=(k_{z},k_{x})$ is the exciton wave number with $%
k_{x}(k_{z}) $ normal (parallel) to the channel direction and $n,l$ is the
quantum number for internal structure of the exciton. Here, $k_{x}$ takes
the value $k_{x}=2\pi n_{x}/Nd,$ with $n_{x}$ an integer that is limited to
one Brillouin zone($n_{x}=1,2,...,N-1$)\cite{14,19}. The Wannier exciton
state can be expressed as

\begin{equation}
\left| {\bf k},n,l\right\rangle =\sum_{l,\rho ,I,J}U_{{\bf k}nl}^{*}(l,\rho
,I,J)\left| c,(l+\rho ,I+J);v,(l,J)\right\rangle ,
\end{equation}
and the interaction matrix elements can be written as:

\begin{equation}
\left\langle {\bf k},n,l\right| H^{\prime }\left| G\right\rangle
=\sum_{l,\rho ,I,J}\left\langle c,(l+\rho ,I+J);v,(l,J)\right| U_{{\bf k}%
nl}^{*}(l,\rho ,I,J)H^{\prime }\left| G\right\rangle ,
\end{equation}
in which the excited state $\left| c,(l+\rho ,I+J);v,(l,J)\right\rangle $ is
defined as

\begin{equation}
\left| c,(l+\rho ,I+J);v,(l,J)\right\rangle =a_{c,(l+\rho ,I+J)}^{\dagger
}a_{v,(l,J)}\left| G\right\rangle ,
\end{equation}
where $a_{c,(l+\rho ,I+J)}^{\dagger }[a_{v,(l,J)}]$ is the creation
[destruction] operator for an electron in the conduction [valence] band at
site $(l+\rho ,I+J)[(l,J)],$ and $l$ and $\rho $ [$I$ and $J$] are the
vectors parallel [normal] to the channel direction. The expansion
coefficient $U_{{\bf k}nl}^{\ast }(l,\rho ,I,J)$ is the exciton wavefunction
in the quantum channel:

\begin{equation}
U_{{\bf k}nl}^{*}(l,\rho ,I,J)=\frac{1}{\sqrt{N^{\prime }}}\frac{1}{\sqrt{N}}%
\exp (ik_{z}r_{c}+ik_{x}r_{c}^{\prime })F_{nl}(\rho ,x),
\end{equation}
where $r_{c}^{\prime }=\frac{m_{e}^{*}(I+J)+m_{h}^{*}I}{m_{e}^{*}+m_{h}^{*}}%
, $ $r_{c}$ is the center of mass of the exciton in the channel direction,
and $F_{nl}(\rho ,x)$ is the hydrogenic wavefunction in the quantum channel.
Following the derivation in above, one can evaluate the decay rate
straightforwardly:

\begin{eqnarray}
\gamma _{{\bf k}nl} &=&\frac{2\pi e^{2}\hbar }{m^{2}cv}\left| {\bf \epsilon }%
_{{\bf k}\lambda }\cdot {\bf A}_{{\bf k}nl}\right| ^{2}\sum_{k_{y}^{\prime }}%
\frac{1}{N}\frac{\sin ^{2}(\pi n_{y}-Ndk_{y}^{\prime }/2)}{\sin ^{2}(\pi
n_{y}/N-dk_{y}^{\prime }/2)}\times  \nonumber \\
&&(\sum_{k_{x}^{\prime }}\frac{\delta (\frac{E_{{\bf k}nl}}{\hbar }-c\sqrt{%
k_{x}^{\prime 2}+k_{y}^{\prime 2}+k_{z}^{2}})}{\sqrt{k_{x}^{\prime
2}+k_{y}^{\prime 2}+k_{z}^{2}}}),
\end{eqnarray}
where $E_{{\bf k}nl}$ is the exciton energy in the quasi-1D channel.

For convenience, the analysis of equation (23) is addressed to the Frenkel
exciton ($F_{nl}(0,0)$ is equal to unity\cite{10}). Generalization to the
case of Wannier exciton is straightforward. In the $n_{y}=0$ mode, there is
an analytical solution for $N=2$ and is given by

\begin{equation}
\gamma _{{\bf k}nl}\varpropto \frac{\gamma _{0}}{k_{0}d}[1+J_{0}(d\sqrt{E_{%
{\bf k}nl}^{2}/c^{2}\hbar ^{2}-k_{z}^{2}})],
\end{equation}
where $J_{0}$ is the Bessel function of the zeroth order. For $N=3$, there
is also an analytical solution:

\begin{equation}
\gamma _{{\bf k}nl}\varpropto \frac{\gamma _{0}}{k_{0}d}[3+2J_{0}(2d\sqrt{E_{%
{\bf k}nl}^{2}/c^{2}\hbar ^{2}-k_{z}^{2}})+4J_{0}(d\sqrt{E_{{\bf k}%
nl}^{2}/c^{2}\hbar ^{2}-k_{z}^{2}})].
\end{equation}
As $N\rightarrow \infty ,$ the system becomes a crystal film, and the decay
rate can be written as

\begin{equation}
\gamma _{{\bf k}nl}\varpropto \frac{\gamma _{0}}{(k_{0}d)^{2}}\frac{1}{\sqrt{%
E_{{\bf k}nl}^{2}/c^{2}\hbar ^{2}-k^{2}}},
\end{equation}
where ${\bf k}$ is the wave vector of the exciton in the crystal film. In
Fig. 1 we have plotted the decay rate as a function of $N.$ In plotting the
figure, we have assumed $k_{0}=2\pi /\lambda ,$ $\lambda =8000\stackrel{%
\circ }{A},$ and lattice spacing $d=5\stackrel{\circ }{A}$ in the numerical
calculation$.$ With the increasing of channel width, the decay rate shows
oscillatory behavior and approach 2D limit. The origin of the oscillation
behavior can be seen more clearly by adding perfectly reflecting mirrors
(microcavity with thickness $L_{c}$) above and below the quantum channel. If
the mirror plane is parallel to the channel plane, it means the exciton can
only couple to discrete photon modes ($k_{x}^{\prime }=\frac{2\pi }{L_{c}}%
n_{c}$, where $n_{c}$ is integers) in the perpendicular direction.
Considering only the lowest mode $k_{x}^{\prime }=\frac{2\pi }{L_{c}}$ , the
decay rate can be evaluated as

\begin{equation}
\gamma _{{\bf k}nl}=G_{Nk_{z}n_{y}}\frac{2\pi e^{2}\hbar }{m^{2}c^{2}v}\frac{%
\left| {\bf \epsilon }_{{\bf k}\lambda }\cdot {\bf A}_{{\bf k}nl}\right| ^{2}%
}{\sqrt{k_{0}^{2}-(\frac{2\pi }{L_{c}})^{2}-k_{z}^{2}}},
\end{equation}
where the oscillation factor

\begin{equation}
G_{Nk_{z}n_{y}}=\frac{1}{N}\frac{\sin ^{2}(\pi n_{y}-\frac{Nd}{2}\sqrt{%
k_{0}^{2}-(\frac{2\pi }{L_{c}})^{2}-k_{z}^{2}})}{\sin ^{2}(\pi n_{y}/N-\frac{%
d}{2}\sqrt{k_{0}^{2}-(\frac{2\pi }{L_{c}})^{2}-k_{z}^{2}})}
\end{equation}
is similar to the quantum well result\cite{14,21} which comes from the
interference between the radiation fields. As can be seen from equation
(27), the exciton modes with $k_{0}<q=\sqrt{(\frac{2\pi }{L_{c}}%
)^{2}+k_{z}^{2}}$ have vanishing decay rate. These exciton modes do not
radiate at all and photon trapping occurs. These dark modes also occur in a
2D thin film. However, it is hard to examine them directly because of the
randomness of $q$ in a thin film. With the recent developments of
fabrication technology, it is now possible to fabricate the planar
microcavities incorporating quantum wires\cite{19}. If the thickness $L_{c}$
is equal to the wavelength of the photon emitted by bare exciton(without
external field), one can examine the dark mode directly by changing $k_{0}$
with external field.

One also notes that as the value of $k_{0}$ is equal to $\sqrt{(\frac{2\pi }{%
L_{c}})^{2}+k_{z}^{2}}$ (resonant mode), the decay rate goes to infinity. In
the work of Ref.[19], C. Constantin {\em et al}. investigated the transition
from nonresonant mode to resonant coupling between quantum confined
one-dimensional carriers and two-dimensional photons states in a
wavelength-long planar Bragg microcavity incorporating strained In$_{0.15}$Ga%
$_{0.85}$/GaAs V-groove quantum wires. They found when the excitonic
transition energy is resonant with the cavity mode, the emission rate into
this mode is significantly enhanced. This significant feature is just the
singularity in equation (27) and can be explained easily by present model.

In summary, we have calculated the decay rate of the exciton in a quasi-1D
channel with channel width $L$. For small channel width ($N=$1,2, and 3),
analytical solutions can be evaluated straightforwardly. Similar to the case
of quantum well system, the decay rate of the exciton shows oscillatory
behavior with the increasing of channel width $L.$ As $L\rightarrow \infty ,$
the decay rate approaches 2D limit correctly. Second, when the quasi 1-D
channel is incorporated with planar microcavities, it becomes possible to
examine the dark modes of the exciton. The distinguishing features are
pointed out and may be observable in a suitably designed experiment.

This work is supported partially by the National Science Council, Taiwan
under the grant number NSC 90-2112-M-009-018.

\newpage

\subsection{FIGURE CAPTION}

Fig. 1 Decay Rate of the superradiant exciton as a function of $N$ $($%
channel width $L=Nd).$ The vertical and horizontal units are $\gamma
_{0}/(k_{0}d)$ and $N,$ respectively.

\end{document}